\documentclass[twocolumn]{revtex4-1} 

\usepackage{graphicx}
\usepackage[export]{adjustbox}
\usepackage{amsmath}
\usepackage{subfig}
\usepackage[justification=justified]{caption}
\usepackage{verbatim}
\usepackage{color}
\usepackage{pdfpages}

\newcommand{\rr}{\mathbf{r}}

\newcommand{\qpar}{\mathbf{q}_\parallel}

\newcommand{\rpar}{\mathbf{r}_\parallel}
\newcommand{\mr}{\mathrm}

\usepackage[abs]{overpic}

\begin{document}
\title{The dielectric genome of van der Waals heterostructures}

\author{Kirsten Andersen}
\email{kiran@fysik.dtu.dk}
\affiliation{Center for Atomic-scale Materials Design, Department of
Physics, Technical University of Denmark, DK - 2800 Kgs. Lyngby, Denmark}

\author{Simone Latini}
\affiliation{Center for Atomic-scale Materials Design, Department of Physics, Technical University of Denmark, DK - 2800 Kgs. Lyngby, Denmark}
\affiliation{Center for Nanostructured Graphene, Technical University of Denmark, DK - 2800 Kgs. Lyngby, Denmark}

\author{Kristian S. Thygesen}
\email{thygesen@fysik.dtu.dk}

\affiliation{Center for Atomic-scale Materials Design, Department of Physics, Technical University of Denmark, DK - 2800 Kgs. Lyngby, Denmark}
\affiliation{Center for Nanostructured Graphene, Technical University of Denmark, DK - 2800 Kgs. Lyngby, Denmark}

\keywords{van der Waals heterostructures, 2D materials, density-functional theory, dielectric function, excitons, plasmons}

\begin{abstract}
Vertical stacking of two-dimensional (2D) crystals, such as graphene and hexagonal boron nitride, has recently lead to a new class of materials known as van der Waals heterostructures (vdWHs) with unique and highly tunable electronic properties. Ab-initio calculations should in principle provide a powerful tool for modeling and guiding the design of vdWHs, but in their traditional, form such calculations are only feasible for commensurable structures with a few layers. Here we show that the dielectric properties of realistic, incommensurable vdWHs comprising hundreds of layers can be calculated with ab-initio accuracy using a multi-scale approach where the dielectric functions of the individual layers (the dielectric building blocks) are coupled simply via their long-range Coulomb interaction. We use the method to illustrate the 2D-3D dielectric transition in multi-layer MoS$_2$ crystals, the hybridization of quantum plasmons in large graphene/hBN heterostructures, and to demonstrate the intricate effect of substrate screening on the non-Rydberg exciton series in supported WS$_2$.
\end{abstract}
\maketitle

The class of 2D materials which started with graphene is rapidly expanding and now includes metallic and semiconducting transition metal dichalcogenides\cite{Wang2012} in addition to group III-V semi-metals, semiconductors and insulators\cite{Sahin2009}. These atomically thin materials exhibit unique opto-electronic properties with high technological potential\cite{Britnell2013,Choi2013, Shih2014,Woessner2014,Withers2015}. However, the 2D materials only form the basis of a new and much larger class of materials consisting of vertically stacked 2D crystals held together by weak van der Waals forces. In contrast to conventional heterostructures which require complex and expensive crystal-growth techniques to epitaxially grow the single-crystalline semiconductor layers, vdWHs can be stacked in ambient conditions with no requirements of lattice matching. The latter implies a weaker constraint, if any, on the choice of materials that can be combined into vdWHs.

\begin{figure*}
  \includegraphics[width = 0.9\linewidth]{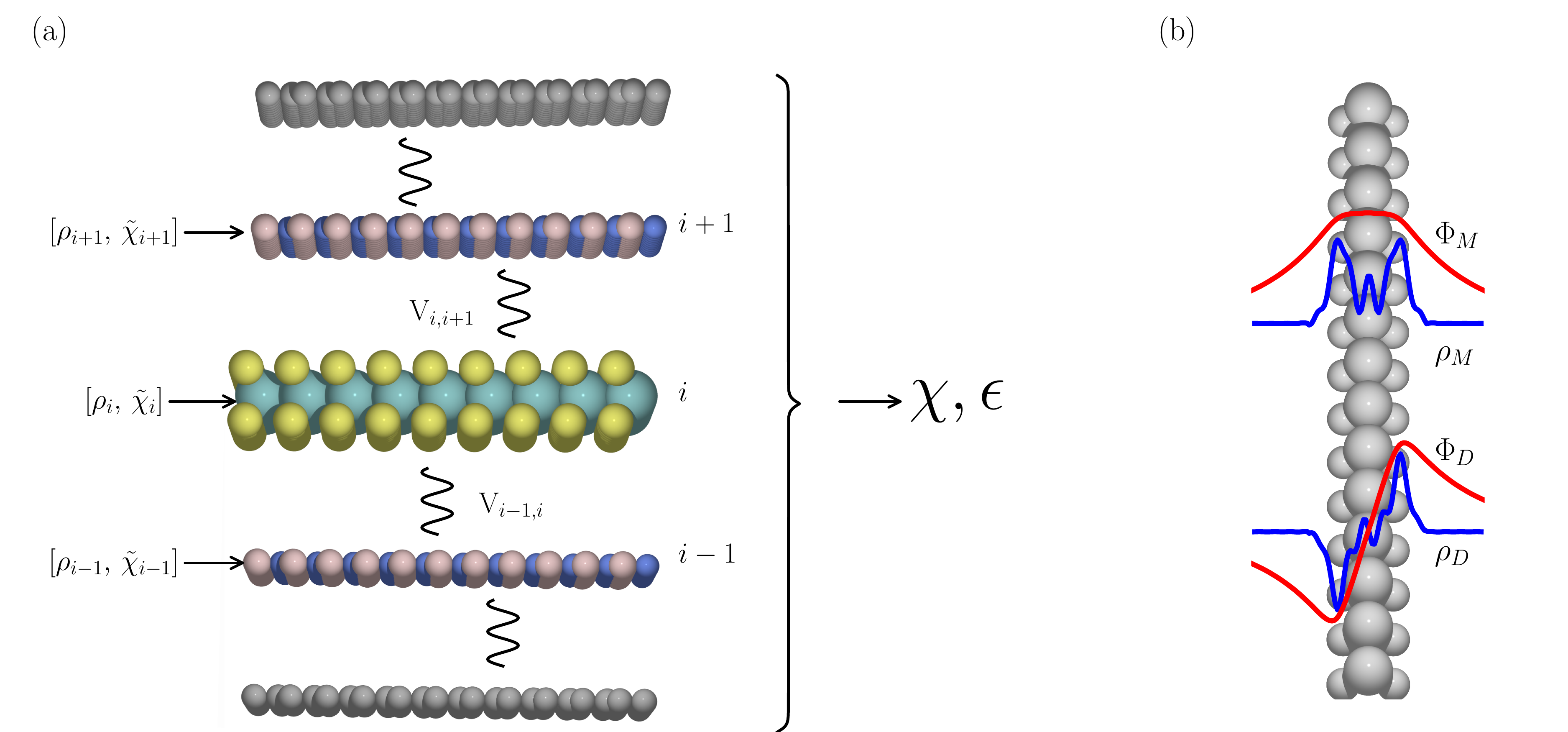}
   \caption{Schematic of the QEH model. (a) The density response function and dielectric function of the heterostructure are calculated from the dielectric building blocks of the individual layers assuming a purely electrostatic interaction between the layers. The dielectric building blocks are calculated ab-initio for the isolated layers. They comprise the monopole and dipole components of the density response function, $\tilde \chi_{M/D}$, together with the spatial shape of the electron density, $\rho_{M/D}(z)$, induced by a constant and linear applied potential, respectively. (b) Monopole and dipole induced densities (blue) together with the associated potentials (red) for monolayer MoS$_2$.}
 \label{scheme}
\end{figure*}

The weak inter-layer binding suggests that the individual layers of a vdWH largely preserve their original 2D properties modified only by the long range Coulomb interaction with the surrounding layers. Turning this argument around, it should be possible to predict the overall properties of a vdWH from the properties of the individual layers. In this Letter we show that this can indeed be achieved for the dielectric properties. Conceptually, this extends the Lego brick picture used by Geim and Grigorieva\cite{Geim2013} for the atomic structure of a vdWH, to its dielectric properties. Specifically, we develop a semi-classical model which takes as input the dielectric functions of the individual isolated layers computed fully quantum mechanically and condensed into the simplest possible representation, and couple them together via the Coulomb interaction, see Figure \ref{scheme}. Despite the complete neglect of interlayer hybridization, the model provides an excellent account of both the spatial and dynamical dielectric properties of vdWHs. The condensed representation of the dielectric function of all isolated 2D crystals can thus be regarded as the dielectric genome of vdWHs. 

In addition to its conceptual value, our approach overcomes a practical limitation of conventional first-principles methods. Such methods are not only computationally demanding, but also rely on periodic boundary conditions which are incompatible with the incommensurable interfaces found in vdWHs. In fact, for many purposes, an in-plane lattice mismatch between neighbouring 2D crystals is preferred because it reduces the interlayer coupling and thus minimises the risk of commensurate-incommensurate transitions\cite{Woods2014}, and formation of Moire patterns\cite{Kang2013} and associated band structure reconstructions\cite{Lu2014} which are typical for systems with similar lattice constants. This emphasises the need for alternative approaches for modelling vdWHs.


The dielectric function is one of the most important material response functions. It determines the effective interaction between charged particles in the material, contains information about the collective oscillations of the electron gas (plasmons)\cite{Pitarke2007}, and enters as a fundamental ingredient in many-body calculations of e.g. excitons and quasiparticle band structures\cite{Onida2002,Hybertsen1986}. 

The (inverse) dielectric function is related to the electron density response function, $\chi$, via 
\begin{equation}\label{eq:epsilon}
\epsilon^{-1}(\rr,\rr', \omega) = \delta(\rr-\rr')+\int \frac{1}{\vert \rr-\rr''\vert } \chi(\rr'', \rr', \omega)d\rr''.
\end{equation}
In our quantum-electrostatic heterostructure (QEH) model the calculation of the dielectric function is divided into two parts. In the first part the in-plane averaged density response function of each of the freestanding layers, $\chi_i(z, z', \qpar, \omega)$, are obtained from ab-initio calculations. In practice we treat the in-plane momentum transfer, $\qpar$, as a scalar since  most 2D materials are isotropic within the plane. From $\chi_i$ we calculate the magnitude of the monopole/dipole component of the density induced by a potential with a constant/linear variation across the layer and in-plane variation $\exp(i \rpar \cdot \qpar)$:
\begin{align}\label{eq:chi}
&\tilde \chi_{i\alpha}(\qpar, \omega) = \int z^\alpha \chi_i(z, z', \qpar, \omega) z'^\alpha dz dz'.
\end{align}
Here $\alpha=0,1$ for the monopole and dipole components, respectively. In addition we calculate the spatial form of the induced density, $\rho_{i\alpha}(z,\qpar)$. With a proper normalization of $\rho_{i\alpha}$ we can then write 
\begin{align}\label{eq:rho}
\int \chi_i(z, z',\qpar,\omega) z'^\alpha dz' = \tilde{\chi}_{i\alpha}(\qpar,\omega) \rho_{i\alpha}(z,\qpar)
\end{align}
We have found that while $\tilde \chi_{i\alpha}$ depends strongly on frequency, $\rho_{i\alpha}$ does not. The data set $(\tilde \chi_{i\alpha},\rho_{i\alpha})$ with $\alpha=0,1$ or equivalently $\alpha=M,D$ constitutes the dielectric building block of layer $i$, as illustrated in Figure \ref{scheme}. According to Eq. (\ref{eq:rho}) the dielectric building block allows us to obtain the density induced in the (isolated) layer $i$ by a constant/linear potential. It is straightforward to extend the dielectric building blocks to account for higher-order moments in the induced density described by $\alpha>1$, but we have found the dipole approximation to be sufficient in all cases considered.

In the second part of the QEH model, the density response function of the vdWH in the discrete monopole/dipole representation is obtained by solving a Dyson equation that couples the dielectric building blocks together via the Coulomb interaction. The Dyson equation for the full density response function giving the magnitude of the monopole/dipole density on layer $i$ induced by a constant/linear potential applied to layer $j$, reads (omitting the $\qpar$ and $\omega$ variables for simplicity)
\begin{align}\label{eq:dysonhet}
\chi_{i\alpha, j\beta}& =
\tilde \chi_{i\alpha} \delta_{i\alpha,j\beta} \: + \tilde \chi_{i\alpha} \sum_{k\neq i ,\gamma} V_{i\alpha,k\gamma}  \, \chi_{k \gamma, j\beta}.
\end{align}
The Coulomb matrices are defined as
\begin{align}\label{eq:coulomb}
V_{i\alpha,k\gamma} (\qpar)=\int \rho_{i\alpha}(z,\qpar) \Phi_{k\gamma}(z,\qpar) dz
\end{align}
where $\Phi_{k\gamma}$ is the potential associated with the induced density, $\rho_{k\gamma}$, which we calculate on a uniform grid by solving a 1D Poisson equation. Note that we leave out the self-interaction terms in Eq. (\ref{eq:dysonhet}) since the intralayer Coulomb interaction is already accounted for by the uncoupled $\tilde \chi_{i\alpha}$. The (inverse) dielectric function of Eq. (\ref{eq:epsilon}) in the monopole/dipole basis becomes
\begin{align}\label{eq:eps}
\epsilon^{-1}_{i\alpha, j\beta}&(\qpar,\omega) =
\delta_{i\alpha,j\beta}  +  \sum_{k \gamma} V_{i\alpha,k\gamma} (\qpar) \, \chi_{k \gamma, j\beta}(\qpar,\omega).
\end{align}
More details on the method and computations are provided in the supporting information.

\begin{figure}
  \includegraphics[width = \linewidth]{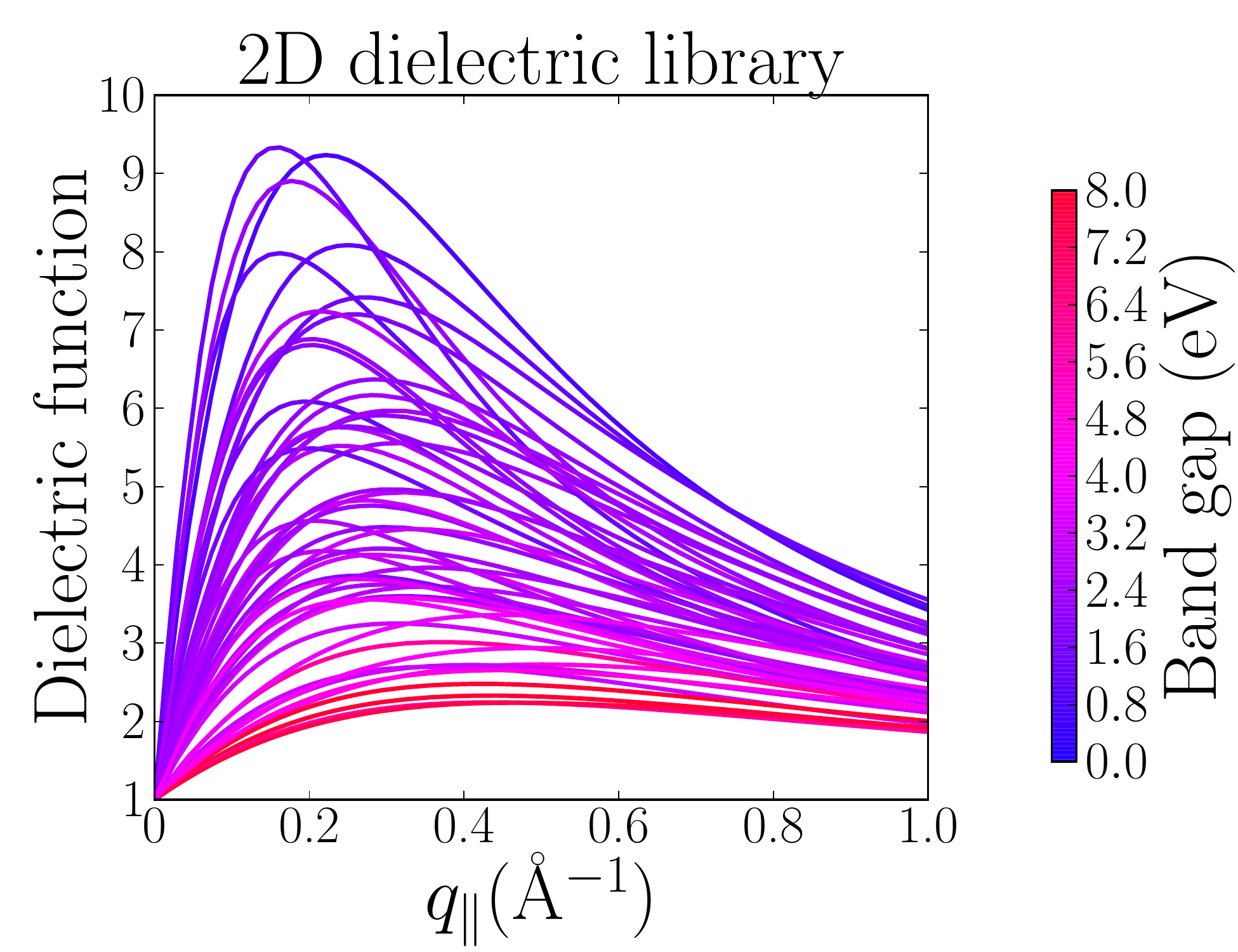}
   \caption{The static dielectric function $\epsilon(\qpar,\omega=0)$ of the 51 transition metal dichalcogenides and oxides included in the database, that span a large range of magnitudes. The relation to the quasiparticle G$_0$W$_0$ band gap of the materials, calculated in Ref.~\cite{rasmussen_computational_2015} is shown in color.}
 \label{database}
\end{figure}

A database containing the dielectric building blocks of a large collection of 2D materials has been constructed, and is available from our website \cite{database}. It presently contains more than 50 transition metal dichalcogenides and oxides, graphene at different doping levels, and hBN, and more materials are being added. From here the data files can be downloaded together with a Python module for calculating the dielectric function and associated properties of any combination of these materials. QEH model calculations for vdWHs containing a few hundred layers can be performed on a standard PC. In Figure \ref{database} we show the $\qpar$-dependent static dielectric functions of the monolayer transition metal dichalcogenides and -oxides presently contained in our database (for a complete overview of the materials see Ref.~\cite{rasmussen_computational_2015}).   All the dielectric functions show the same qualitative form, in particular they become 1 for $\qpar\to0$ and $\qpar \to \infty$, however there is quite some variation in their magnitude. As expected the size of the dielectric function correlates well with the size of the band gap of the material indicated by the colour. 

First-principles calculations were performed with the GPAW code\cite{GPAW2,Yan2011a}. Single-particle wave functions and energies were calculated within the local density approximation (LDA) using 400 eV plane wave cut-off and at least $45\times45$ sampling of the 2D Brillouin zone. Density response functions and dielectric functions were calculated within the random phase approximation (RPA). The RPA does not include electron-hole interactions, but generally yields good results for the static dielectric properties of semi-conductors and dynamical response of metals. Except for MoS$_2$ bulk, we included at least 15 \AA~of vacuum in the super cells perpendicular to the layers and applied a truncated Coulomb kernel to avoid long range screening between periodically repeated structures. All response functions were calculated in a plane wave basis including reciprocal lattice vectors up to at least 50 eV. A similar cut off was used for the sum over empty states and convergence was carefully checked. The frequency dependence of the response functions was represented on a non-linear frequency grid ranging from 0 to 35 eV, with an initial grid spacing of 0.02 eV. All details of the calculations and atomic structure geometries are provided in the supporting information.

 \begin{figure*}
\includegraphics[width = \linewidth]{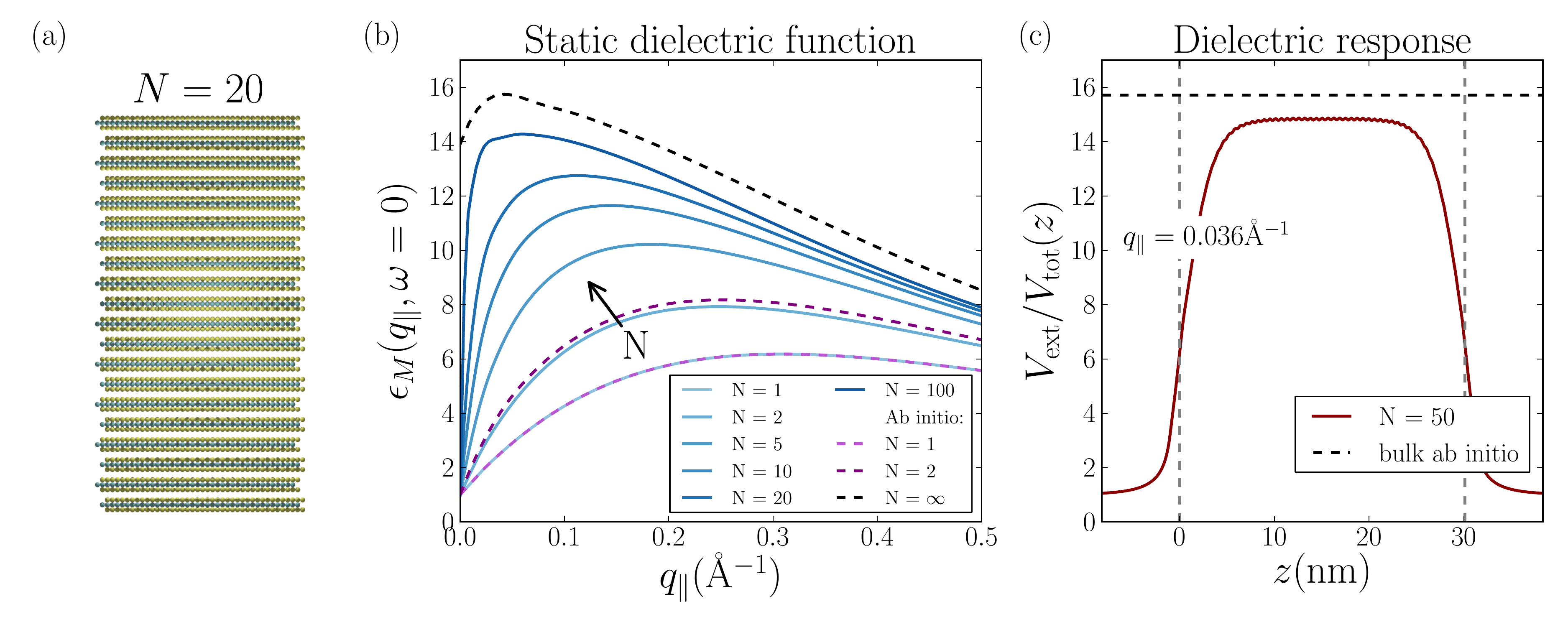}
  	  \caption{2D-3D transition of the dielectric function. (a) Atomic structure of a 20 layer MoS$_2$ slab. (b) The macroscopic static dielectric function $\epsilon_M(\qpar, \omega = 0)$ as a function of the in-plane momentum transfer for different number of layers, $N$. The macroscopic dielectric function relates the total potential averaged over the width of the slab to an external potential of the form $V_{ext}(r_\Vert,z)=\exp(ir_\Vert \qpar)$. The dielectric functions increase monotonically with $N$ converging slowly towards the dielectric function of bulk MoS$_2$ obtained from an ab-initio calculation. Excellent agreement between the QEH model and the ab initio results are seen for $N=1,2$. The slow convergence towards the bulk result is due to the strong spatial variation of the induced potential in the surface region of the slabs. This can be seen in panel (c) which shows $V_{ext}/V_{tot}(z)$, i.e.~the local dielectric function, for an external potential constant across the slab and with in-plane wave vector $\qpar=0.036$\AA$^{-1}$ for $N=50$.}
  \label{mos2macro}
\end{figure*}

As a first application of the QEH model, we study how the (static) dielectric function of a 2D material evolves as the layer thickness increases towards the bulk. One of the most characteristic differences between 2D and 3D materials is the behaviour of the dielectric function in the long wave length limit: For a bulk semiconductor, the dielectric function $\epsilon(\qpar)$ tends smoothly to a constant value larger than unity as $\qpar \to 0$. In contrast $\epsilon(\qpar)=1+O(\qpar)$ for a 2D semiconductor implying a complete absence of screening in the long wave length limit \cite{Cudazzo2011,Huser2013}. 

Ab initio calculations were performed for the dielectric function of MoS$_2$ monolayer, bilayer, and bulk, and the QEH model was used for multilayer structures up to 100 layers. Figure \ref{mos2macro} (b) shows the dielectric functions averaged over the slabs, i.e.~the macroscopic dielectric function, as function of the in-plane momentum transfer. For large $\qpar$ the dielectric functions show similar behavior. However, whereas $\epsilon(0)=14$ for the bulk, the dielectric functions of the slabs decrease sharply to 1 for small $\qpar$. This demonstrates that the dielectric properties of a vdWH of thickness $L$ are 2D like for $\qpar \ll 1/L$ and 3D like for $\qpar \gg 1/L$. Interestingly, also the result for bulk MoS$_2$ shows reminiscence of the 2D nature of the constituent layers, where the magnitude of the dielectric function has a slight drop when $\qpar \to 0$.

The QEH model describes the change in the dielectric function from mono- to bilayer very accurately in spite of the well known differences between the mono- and bilayer band structures\cite{Cheiwchanchamnangij2012}. This shows that hybridisation driven band structure effects, i.e.~quantum confinement, have negligible influence on the dielectric properties of a vdWH and is the main reason for the success of the QEH model. The model result seems to converge towards the ab initio bulk result, however, convergence is not fully reached even for $N=100$. The slow convergence towards the bulk result is mainly due to the spatial variation of the induced potential across the slab. In Figure \ref{mos2macro} (c) we show the $z$-dependent dielectric function defined as $\epsilon(z)=V_{ext}/V_{tot}(z)$, for a constant (along $z$) external potential with a long wavelength in-plane variation for $N=50$. Although $\epsilon(z)$ is close to the ab-initio bulk value (dashed line) in the middle of the slab, screening is strongly suppressed in the surface region. Increasing the slab thickness beyond 50 layers brings the QEH result even closer to the bulk result in the middle of the slab, but a small underestimation remains originating from the difference in the band structures of the monolayer and bulk systems. The suppressed screening in the surface region is a direct consequence of the anisotropic nature of the layered MoS$_2$ crystals which limits the screening of perpendicular fields relative to in-plane fields, and is expected to be a general property of vdWHs.

The model can also be used to calculate the response to fields polarized along the $z$-direction, i.e. perpendicular to the layers. In this case the perpendicular component, $\epsilon_{zz}(q_z=0)$, can be calculated by applying an external potential with a linear variation along $z$. In the discrete basis of the QEH model, such a field is represented by a vector with 0 for all monopole components and 1 for all dipole components. Comparing the averaged slope of the total potential to the slope of the applied linear potential for a slab of N=100 layers of MoS2 yields $\epsilon_{zz}= 7.8$.  This value is somewhat larger than the bulk value of 6.03, however, due to long range surface effects it is not necessarily to be expected that the two numbers should coincide. In fact, we find excellent agreement between the QEH model and full ab-initio calculation of $\epsilon_{zz}$ for a four layer MoS$_2$ slab (see supporting information).

\begin{figure*}
 \includegraphics[width = \linewidth]{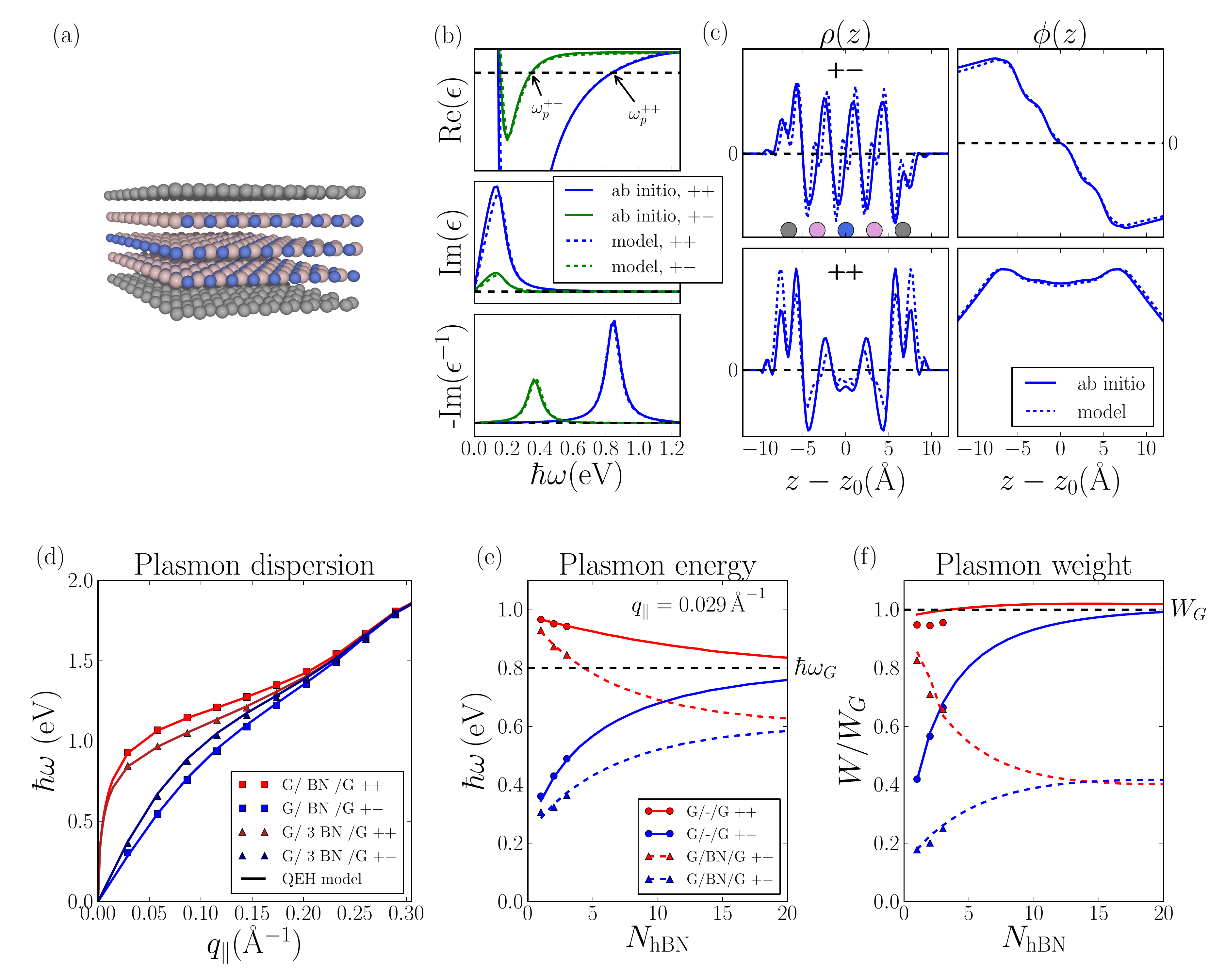}
  \caption{Plasmons in graphene/hBN heterostructures. (a) Two graphene sheets separated by three layers of hBN.  (b) Eigenvalues of the heterostructure dielectric function $\epsilon(\omega)$. Only the two eigenvalue curves that fullfill the plasmon condition $\mr{Re}\epsilon_n(\omega_P)=0$ are shown. (c) The eigen-potential, $\psi(\omega_P)$, and associated density, $\rho(\omega_P)$, of the plasmon modes. The plasmons  correspond to the antisymmetric ($+-$) and symmetric ($++$) combinations of the isolated graphene plasmons. (d) Plasmon dispersion for heterostructures containing 1 and 3 layers of hBN. Full lines denote the QEH model while ab-initio results are denoted by symbols. (e+f) Energy and weight of the plasmon modes for up to 20 layers hBN between the graphene sheets. Results for equivalent structures with vacuum filling the gap are also shown. Dashed black lines indicate the plasmon energy and weight in an isolated graphene sheet. Overall, the QEH model is in excellent agreement with the full ab initio calculations performed for up to 3 layers hBN.}
  \label{graphene-BN}
\end{figure*}

Next, we consider the hybridisation of plasmons in graphene sheets separated by hBN, see Figure \ref{graphene-BN}(a). Plasmons in graphene on hBN were recently found to propagate with low loss \cite{Woessner2014}, and the close to perfect lattice match between the two layers enables full ab initio calculations for the thinnest heterostructures. Here we use doped graphene that has a finite density of states at the Fermi level, giving rise to two-dimensional plasmons with energies in the regime 0-2 eV. The plasmon energies goes to zero in the optical limit, $\qpar \rightarrow 0$ as is characteristic for plasmons in 2D metals\cite{Hwang2007, Shin2011}. We calculate the effect of hBN on the plasmons using the QEH model for up to 20 layers of hBN and compare to full ab-initio calculations for 1-3 layers of hBN.

To identify the plasmons of the heterostructure we follow Ref. \cite{Andersen2012}. In brief, we compute the eigenvalues, $\epsilon_n(\omega)$, of the heterostructure dielectric function for each frequency point and identify a plasmon energy, $\hbar \omega_P$, from the condition $\text{Re}\epsilon_n(\omega_P)=0$, see Figure \ref{graphene-BN}(b). The corresponding eigenvector, $\phi_n(\omega_P)$, represents the potential associated with the plasmon oscillation, see panel (c). This analysis identifies two plasmons corresponding to the symmetric ($++$) and antisymmetric ($+-$) combinations of the graphene plasmons as previously found for two freestanding graphene sheets \cite{Hwang2009}. For 1-3 hBN layers, the QEH model perfectly reproduces the ab-initio results for the dielectric eigenvalues, plasmon energy, and weight. The latter was defined as the area under the peaks in the loss function $-\mr{Im}\epsilon^{-1}(\qpar,\omega)$, see panel (b). The densities and potentials of the plasmon eigenmodes shown in panel (c) are also reproduced fairly accurately by the model, where the qualitative differences for the induced densities, $\rho(z)$, are due to the use of a limited basis of the monopole and dipole response for each layer. In panels (e-f) the result of full ab-initio calculations are shown by symbols while the QEH results are shown by continuous lines. The effect of the hBN buffer (dashed lines) is to red shift and damp the plasmons compared to the result for two graphene sheets separated by the same amount of vacuum (full lines). This is also reflected by the relatively large amount of electron density located on the hBN during the plasma oscillation, see panel (c).

\begin{figure*}
  \includegraphics[width = \linewidth]{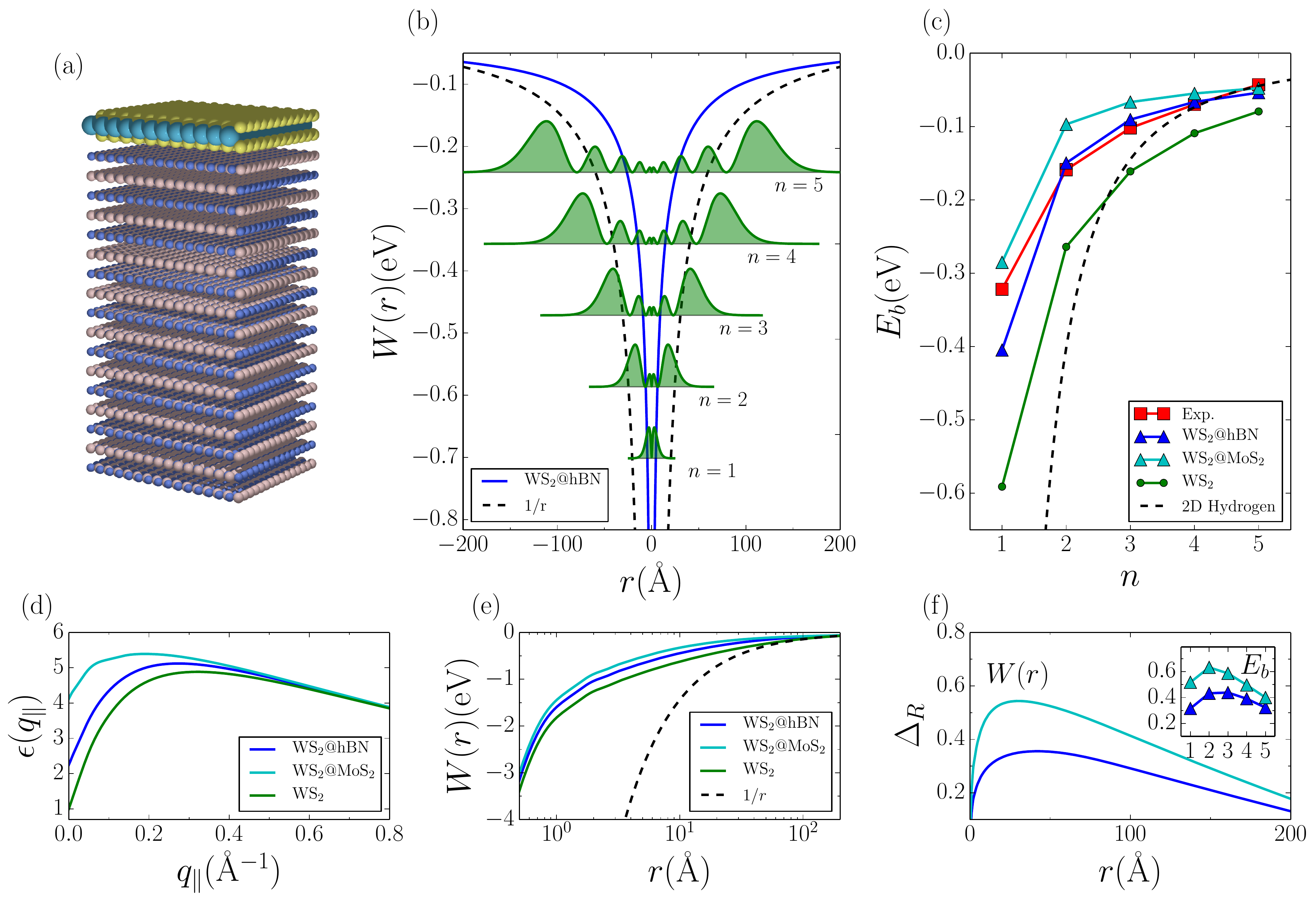}
  \caption{Excitons in supported WS$_2$. (a) Monolayer WS$_2$ adsorbed on a h-BN substrate. (b) The screened interaction between an electron and a hole localised within a WS$_2$ monolayer adsorbed on hBN. For comparison the unscreened $1/r$ potential is shown. The radial probability distribution of the first five excitons, $r\vert F(r)\vert^2$, are also shown (arbitrary normalization). (c) The calculated binding energies of the lowest five excitons in freestanding WS$_2$ (green dashed) and WS$_2$ on hBN (blue) and MoS$_2$ (cyan). Experimental values from Ref. \cite{Chernikov2014} for WS$_2$ on SiO$_2$ are shown in red. The 2D hydrogen model with a $1/\epsilon r$ potential is shown for $\epsilon=1.7$. (d) The dielectric function of the WS$_2$ layer defined as $\epsilon(q)=V(q)/W(q)$, where $V(q)$ and $W(q)$ are the bare and screened interaction in the WS$_2$ layer, respectively. (e) The screened interaction in the WS$_2$ layer as function of $\log(r)$. (f) The relative difference between the screened interaction in the supported and freestanding WS$_2$. Inset shows the relative difference between $E_b$ for the supported and freestanding WS$_2$.}
  \label{fig:excitons}
\end{figure*}

Finally, we explore some characteristic features of excitons in freestanding and supported 2D semiconductors. A straight forward generalisation of the well known Mott-Wannier model\cite{Wannier1937} leads to the following eigenvalue equation for the excitons of a 2D semiconductor\cite{Cudazzo2011, Berkelbach2013}:
\begin{equation}
\left[-\frac{\nabla_{2D}^2}{2\mu_{ex}}+W(\rr)\right]F(\rr)=E_bF(\rr),
\label{eq:MWHamiltonian}
\end{equation}
where $E_b$ is the exciton binding energy, $F(\rr)$ is the wave function, $\mu_{ex}$ is the effective mass, and $W(\rr)$ is the screened electron-hole interaction. Assuming that the electron and hole are localised in layer 1, the Fourier transformed screened electron-hole interaction is obtained from the static ($\omega=0$) response function Eq. (\ref{eq:dysonhet}) and Coulomb interaction matrix Eq. (\ref{eq:coulomb}) of the QEH model,
\begin{equation}
W(\qpar) = V_{1M,1M}+\sum_{i \alpha,j\beta} V_{1 M,j \beta}(\qpar) \chi_{j \beta, i \alpha}(\qpar)V_{i \alpha, 1 M} (\qpar).
\label{eq:W_q}
\end{equation}
The first term is the bare, i.e. unscreened, electron-hole interaction in layer 1 under the assumption that the electron and hole densities can be represented by the induced monopole density, $\rho_{1M}(z)$. The second term describes the screening from the surrounding layers and layer 1 itself. Note that the above equation can be easily generalised to describe the screened interaction between charges localised in different layers (relevant for indirect excitons). 

In Ref. \cite{Chernikov2014} Chernikov et al. observed a peculiar non-hydrogenic Rydberg series for the excitons in a single layer of WS$_2$ adsorbed on a SiO$_2$ substrate. Here we use the QEH model to calculate the screened electron-hole interaction within the WS$_2$ layer from the dielectric function of the full heterostructure. Since the QEH is applicable only to layered materials we place WS$_2$ on a 100 layer thick slab of hBN which has dielectric constant very similar to that of SiO$_2$ (both around 4). For comparison we performed similar calculations using MoS$_2$ as substrate (dielectric constant larger than SiO$_2$). Figure \ref{fig:excitons} (c) shows the five lowest $s$-excitons calculated from Eq. (\ref{eq:MWHamiltonian}) for both freestanding and supported WS$_2$. For freestanding WS$_2$, we obtain $E_b=0.59$ eV for the lowest exciton in good agreement with previous ab-initio calculations\cite{Shi2013}. The enhanced screening from the substrate lowers the exciton binding energies bringing the entire series closer to the experimental values (red), in particular for the hBN substrate.  

The dielectric function of the WS$_2$ layer defined as $\epsilon(q)=V(q)/W(q)$, where $V(q)$ and $W(q)$ are the bare and screened interaction in the WS$_2$ layer, respectively. Figure \ref{fig:excitons} (d) shows that the dielectric function of the supported WS$_2$ layer exceeds unity in the $\qpar\to 0$ limit. For structures of finite width, $L$, the dielectric function will in practice tend to unity for very small $\qpar \ll 1/L$. Here the result have been extrapolated to infinite substrate thickness, where $\epsilon(\qpar)$ tends to a value larger than unity. This means that the nature of the screening within the layer is not strictly 2D because the bulk substrate is able to screen the long wave length fields. In real space, the screened potentials diverge as $\log(r)$ for small $r$ and decay as $1/r$ for large $r$, see panel (e). In panel (f) we show how the substrate affects $W(r)$: The relative deviation from $W(r)$ of the freestanding layer vanishes for small and large $r$ but becomes significant at intermediate distances. As a consequence, the substrate-induced change in the exciton binding energy is relatively larger for intermediate exciton sizes. These results clearly demonstrate the profound, nonlocal influence of substrates on the dielectric screening and excitations in 2D materials.  

In conclusion, we have demonstrated that the spatial and dynamical dielectric properties of a vdWH can be accurately and efficiently obtained from the dielectric properties of its constituent 2D crystals. The presented quantum-electrostatic heterostructure model (QEH) exploits this feature and enables the calculation of the dielectric properties and collective electronic excitations of realistic incommensurable heterostructures with ab-initio precision. The dielectric building blocks for more than fifty different 2D materials are available in an open database allowing 2D materials researchers to efficiently predict and design the dielectric properties of realistic vdWHs.

\section*{acknowledgement}
The authors thank Karsten Jacobsen for inspiring discussions. The authors acknowledge support from the Danish Council for Independent Research's Sapere Aude Program through grant no. 11-1051390. The Center for Nanostructured Graphene (CNG) is sponsored by the Danish National Research Foundation, Project DNRF58.

\section*{Supporting information available}
Detailed description of our quantum-electrostatic heterostructure (QEH) model and the computational details for all the ab-initio calculations are given in the supporting information.

\bibliography{bibtex}

\cleardoublepage
\onecolumngrid
\section*{}


\section*{Supplementary material}

# Supporting information for: The dielectric genome of van der Waals Heterostructures

Kirsten Andersen and Simone Latini
*Center for Atomic-scale Materials Design, Department of Physics*
*Technical University of Denmark, DK - 2800 Kgs. Lyngby, Denmark*

Kristian S. Thygesen*
*Center for Atomic-scale Materials Design, Department of Physics, and Center for Nanostructured Graphene*
*Technical University of Denmark, DK - 2800 Kgs. Lyngby, Denmark*

In this supplementary material we provide a detailed description of our quantum-electrostatic heterostructure (QEH) model including the precise definition of the dielectric building blocks. In addition we detail the spectral analysis used to identify the plasmon eigen modes for the graphene/hBN structures and describe the calculation of the screened electron-hole interaction used in the 2D exciton model. Finally, we provide computational details for all the ab-initio calculations presented in the Letter.

## I. FORMAL MATTERS

Within linear response theory, the induced density due to an external field of the form $V_{ext}(\mathbf{r}, t) = V_{ext}(\mathbf{r}, \omega)e^{i\omega t}$, is described by the density response function, $\chi(\mathbf{r}, \mathbf{r}', \omega)$:

$$n_{ind}(\mathbf{r}, \omega) = \int dr' \chi(\mathbf{r}, \mathbf{r}', \omega) V_{ext}(\mathbf{r}', \omega), \quad (1)$$

The density response function can be obtained from its non-interacting counterpart, $\chi^0(\mathbf{r}, \mathbf{r}', \omega)$, that gives the response to the *total* field, by solving the Dyson equation in the random phase approximation (RPA):

$$\chi(\mathbf{r}, \mathbf{r}', \omega) = \chi^0(\mathbf{r}, \mathbf{r}', \omega) + \int\int d\mathbf{r}_1 d\mathbf{r}_2 \chi^0(\mathbf{r}, \mathbf{r}_1, \omega) \frac{1}{|\mathbf{r}_1 - \mathbf{r}_2|} \chi(\mathbf{r}_2, \mathbf{r}', \omega). \quad (2)$$

For modelling of vdWHs, this equation is favourably split into two parts, namely the intra-layer and inter-layer parts, as described below.

We are assuming a basis set consisting of layer centred functions, $\{\phi_{i\alpha}\}$, where $i$ denotes the layer. Defining the Coulomb matrix as

$$\mathbf{V}_{i\alpha, j\beta} = \int d\mathbf{r} d\mathbf{r}' \phi_{i\alpha}(\mathbf{r}) \frac{1}{|\mathbf{r} - \mathbf{r}'|} \phi_{j\beta}(\mathbf{r}') \quad (3)$$

we can divide the Coulomb interaction into its intra- and interlayer parts: $\mathbf{V} = \tilde{\mathbf{V}} + \mathbf{V}^{\mathbf{I}}$. The Dyson equation 2 can then be separated into the following two matrix equations

$$\tilde{\chi} = \chi^0 + \chi^0 \tilde{\mathbf{V}} \tilde{\chi} \quad (4)$$

$$\chi = \tilde{\chi} + \tilde{\chi} \mathbf{V}^{\mathbf{I}} \chi. \quad (5)$$

To see this, simply insert Eq. 4 into Eq. 5

$$\chi = \chi^0 + \chi^0 \tilde{\mathbf{V}} \tilde{\chi} + \chi^0 \mathbf{V}^{\mathbf{I}} \chi + \chi^0 \tilde{\mathbf{V}} \tilde{\chi} \mathbf{V}^{\mathbf{I}} \chi \quad (6)$$

$$= \chi^0 + \chi^0 \mathbf{V}^{\mathbf{I}} \chi + \chi^0 \tilde{\mathbf{V}} (\tilde{\chi} + \tilde{\chi} \mathbf{V}^{\mathbf{I}} \chi) \quad (7)$$

$$= \chi^0 + \chi^0 \mathbf{V}^{\mathbf{I}} \chi + \chi^0 \tilde{\mathbf{V}} \chi \quad (8)$$

$$= \chi^0 + \chi^0 (\tilde{\mathbf{V}} + \mathbf{V}^{\mathbf{I}}) \chi, \quad (9)$$

which is the original Dyson equation.

At this point no approximations, except for the RPA, have been introduced. In particular, $\chi^0$ in Eq. 4 is the non-interacting response function of the full vdWH. To make progress we make the assumption that the overlap/hybridization between wave functions (not to be confused with the basis functions) on neighbouring layers can be neglected. This allows us to replace $\chi^0$ of the heterostructure by the sum of $\chi_i^0$ for the individual isolated layers. In practice this means that Eq. 4 can be solved for each layer separately.

We calculate $\chi^0$ for the isolated layers within the RPA using single-particle wave functions and energies from density functional theory (DFT) as described in Ref.[1]. The interacting density response function, $\tilde{\chi}$, for the monolayer is obtained by solving the Dyson equation in a plane-wave basis with a 2D truncated Coulomb Kernel, $\tilde{V}_{\mathbf{G}}^{2D}$:

$$\tilde{V}_{\mathbf{G}, G_z}^{2D} = \frac{4\pi}{G^2} [1 - \cos(G_z L/2)]. \quad (10)$$

The use of a truncated Coulomb interaction is essential to avoid interaction between periodically repeated layers[2]. The truncation length is set to half the unit cell height, $L$. In the plane wave basis, the Dyson equation for the density response function, $\tilde{\chi}$, is then written:

$$\tilde{\chi}_{\mathbf{G}, \mathbf{G}'}(\mathbf{q}_{\parallel}, \omega) = \chi_{\mathbf{G}, \mathbf{G}'}^0(\mathbf{q}_{\parallel}, \omega) + \sum_{\mathbf{G}_1} \chi_{\mathbf{G}, \mathbf{G}_1}^0(\mathbf{q}_{\parallel}, \omega) \tilde{V}_{\mathbf{G}_1}^{2D}(\mathbf{q}_{\parallel}) \tilde{\chi}_{\mathbf{G}_1, \mathbf{G}'}(\mathbf{q}_{\parallel}, \omega), \quad (11)$$

where $\mathbf{q}_{\parallel}$ belongs to the 2D Brillouin zone.

## II. QEH MODEL

### A. The dielectric building blocks

We start by defining the density response function for the individual layers, where the macroscopic average is

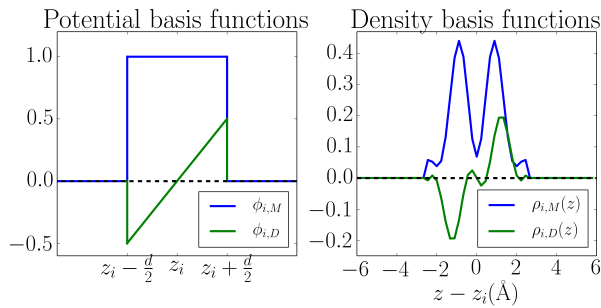

FIG. 1: Basis functions used to represent potentials (left) and induced densities (right) in the QEH model. The example is for graphene at $q_\parallel = 0.029 \text{Å}^{-1}$.

taken in the parallel directions. The response function is then expressed in terms of the perpendicular coordinates $z$ and $z'$, and the magnitude of the momentum transfer parallel to the layer, $q_\parallel$ (we assume isotropic materials, where the response does not depend on the direction of $q_\parallel$, but the method can be straightforwardly generalized to non-isotropic 2D materials):

$$\tilde{\chi}(z, z', q_\parallel, \omega) = \frac{1}{A} \int_A \int_A d\mathbf{r}_\parallel d\mathbf{r}'_\parallel \tilde{\chi}(\mathbf{r}, \mathbf{r}', q_\parallel, \omega)$$
$$= \frac{1}{L} \sum_{\mathbf{G}_z, \mathbf{G}'_z} e^{i\mathbf{G}_z z} \tilde{\chi}_{\mathbf{G}_z, \mathbf{G}'_z}(q_\parallel, \omega) e^{-i\mathbf{G}'_z z'}, \quad (12)$$

where the integration is over the in-plane coordinates, $A$ is the in-plane area of the supercell, and $L$ is the height of the supercell perpendicular to the layer. Integrating over the in-plane coordinates corresponds to taking the zero components $\mathbf{G}_\parallel = \mathbf{G}'_\parallel = 0$ in the plane-wave representation of $\tilde{\chi}_{\mathbf{G},\mathbf{G}'}(\mathbf{q}, \omega)$. Working with $\tilde{\chi}$ instead of $\tilde{\chi}^0$ ensures that local field effects within the isolated layer are exactly taken into account.

For an efficient representation of the response functions and solution of the Dyson equation we need a small yet accurate basis set to represent the induced densities in the layers and the potentials created by these induced densities. To represent potentials we simply use a constant and linear potential corresponding to a first order expansion of the induced potentials, see Fig 1(left). We refer to these as monopole (M) and dipole (D) potentials. The potential basis functions of layer $i$ at position $z_i$ are thus

$$\phi_{i,M}(z) = 1_{[z_i - d/2, z_i + d/2]} \quad (13)$$
$$\phi_{i,D}(z) = (z - z_i) 1_{[z_i - d/2, z_i + d/2]} \quad (14)$$
$$1_C = \begin{cases} 1 & \text{if } z \in C \\ 0 & \text{if } z \notin C \end{cases} \quad (15)$$

where $d$ is a localisation parameter that is set equal to the interplane distance. Since the density response is already confined to the layer, the precise value of $d$ is not essential

and in calculating the matrix elements of the intralayer response function we integrate over all space:

$$\tilde{\chi}_{i\alpha}(q_\parallel, \omega) = \int \int dz dz' \phi_{i,\alpha}(z) \tilde{\chi}(z, z', q_\parallel, \omega) \phi_{i,\alpha}(z') \quad (16)$$
$$\approx \int \int dz dz' (z - z_i)^\alpha \tilde{\chi}_i(z, z', q_\parallel, \omega)(z' - z_i)^\alpha, \quad (17)$$

where $\alpha = \{M, D\}$ or equivalently $\alpha = \{0, 1\}$.

The basis functions can be interpreted as potentials that act on $\chi$. In order to represent the induced densities produced by these potentials, we introduce two density basis functions defined as

$$\rho_{i,\alpha}(z, q_\parallel) = \frac{\int dz' \tilde{\chi}(z, z', q_\parallel, \omega = 0) \phi_{i,\alpha}(z')}{\tilde{\chi}_{i,\alpha}(q_\parallel, \omega = 0)}. \quad (18)$$

As an example, the monopole and dipole density basis functions for monolayer graphene are shown in Fig. 1(right). We have found that the frequency dependence of the basis functions can in general be omitted, while the $q_\parallel$-dependence is not always negligible. Dividing by $\tilde{\chi}_{i,\alpha}(q_\parallel, \omega = 0)$ in Eq. 18 ensures that the density basis function is normalized such that the overlap with the potential basis is unity: $\langle \phi_{i,\alpha} | \rho_{i,\alpha}(q_\parallel) \rangle = 1$, where integration over $z$ is implied. To ease the derivation of the Dyson equation in the monopole/dipole basis, we make the approximation that the potential and density basis functions form a dual basis, i.e.

$$\langle \phi_{i,\alpha} | \rho_{j,\beta}(q_\parallel) \rangle = \delta_{\alpha\beta} \delta_{ij}, \quad (19)$$

where $\alpha, \beta = \{M, D\}$, and $i, j$ are layer indices. This implies that, within the subspace spanned by the basis functions, we have the completeness relation

$$\mathbf{P} = \sum_{i,\alpha} |\rho_{i,\alpha}\rangle \langle \phi_{i,\alpha}| = \hat{1} \quad (20)$$

We note that Eq. (19) is not exact because of the small but finite overlap between potential and density basis functions on neighbouring layers. However, taking this into account gives very small modifications to the resulting vdWH dielectric properties. Finally, we note that working with a dual basis is natural as, in general, the spectral representation of the dielectric function is written in a dual basis of potential and density eigenfunctions[3].

## B. Electrostatic Dyson equation

The Dyson equation (5) for the heterostructure density response function $\chi(z, z', q_\parallel, \omega)$ is now written in the potential basis of dimension $2N \times 2N$, where $N$ is the number of layers. In the following the $(q_\parallel, \omega)$ variables are omitted from the expressions for simplicity. Response functions $\tilde{\chi}$, $\chi$ and Coulomb kernel $V$ are regarded as operators and integration over $\mathbf{r}, \mathbf{r}'$ is implied in the inner

products. The matrix elements of $\chi$ are written in the potential basis:

$$\langle \phi_{i,\alpha} | \chi | \phi_{j,\beta} \rangle = \langle \phi_{i,\alpha} | \tilde{\chi} | \phi_{j,\beta} \rangle + \langle \phi_{i,\alpha} | \tilde{\chi} \, V^I \chi | \phi_{j,\beta} \rangle. \quad (21)$$

The first term on the right hand side is simply the response function of the isolated layers for which we have $\langle \phi_{i,\alpha} | \tilde{\chi} | \phi_{j,\beta} \rangle = \tilde{\chi}_{i,\alpha} \delta_{i\alpha,j\beta}$. In the second term, applying $\langle \phi_{i,\alpha} |$ to $\tilde{\chi}$ returns $\tilde{\chi}_{i,\alpha} \langle \rho_{i,\alpha} |$ (this follows from Eq. 18 and the symmetry of $\tilde{\chi}(z, z')$). Now the completeness relation (20) is inserted between $V^I$ and $\chi$, leading to

$$\langle \phi_{i,\alpha} | \chi | \phi_{j,\beta} \rangle = \tilde{\chi}_{i,\alpha} \delta_{i\alpha,j\beta} + \\ \tilde{\chi}_{i,\alpha} \sum_{k,\alpha'} \langle \rho_{i,\alpha} | V^I | \rho_{k,\alpha'} \rangle \langle \phi_{k,\alpha'} | \chi | \phi_{j,\beta} \rangle$$

This leads to the final Dyson equation for the heterostructure:

$$\chi_{i\alpha,j\beta}(q_\|, \omega) = \tilde{\chi}_{i,\alpha}(q_\|, \omega) \delta_{i\alpha,j\beta} + \\ \tilde{\chi}_{i,\alpha}(q_\|, \omega) \sum_{k,\gamma} V_{i\alpha,k\gamma}(q_\|) \, \chi_{k\gamma,j\beta}(q_\|, \omega). \quad (22)$$

The Coulomb kernel is here defined in the density basis as: $V_{i\alpha,k\alpha'} = \langle \rho_{i,\alpha} | V | \rho_{k,\alpha'} \rangle$. The term $V | \rho_{k,\alpha'} \rangle$ is the potential at $z$ from the density basis function in layer $k$, which is found by solving Poisson's equation for $|\rho_{k,\alpha'}\rangle$ on a real space grid. Since the density parallel to the layer just shows periodic oscillations with wave vector $q_\|$, Poisson's equation reduces to a 1D differential equation:

$$\frac{\partial^2}{\partial z^2} \Phi_{k\alpha'}(z) - q_\|^2 \Phi_{k\alpha'}(z) = -4\pi \rho_{k\alpha'}(z). \quad (23)$$

The elements of the $V$ matrix are then: $V_{i\alpha,k\alpha'} = \langle \rho_{i,\alpha} | \Phi_{k,\alpha'} \rangle$.

## C. The dielectric matrix

The inverse dielectric function is related to $\chi$ through: $\epsilon^{-1} = I - V\chi$. Due to the non-symmetric nature (in $\mathbf{r}$ and $\mathbf{r}'$) of the dielectric function, the elements of $\epsilon^{-1}$ are naturally written using a mixed density/potential basis:

$$\langle \rho_{i,\alpha} | \epsilon^{-1} | \phi_{j,\alpha} \rangle = \delta_{i\alpha,j\beta} + \langle \rho_{i,\alpha} | V \chi | \phi_{j,\beta} \rangle. \quad (24)$$

Upon insertion of the completeness relation (20) this gives

$$\epsilon^{-1}_{i\alpha,j\beta}(q_\|, \omega) = \delta_{i\alpha,j\beta} + \sum_{k,\gamma} V_{i\alpha,k\gamma}(q_\|) \chi_{k\gamma,j\beta}(q_\|, \omega). \quad (25)$$

## III. PLASMONS EIGENMODES

By following a previously developed method for identifying plasmon eigenmodes in nanostructures from ab initio[3], the dielectric matrix for the heterostructure, Eq. 25, is diagonalized to solve the eigenvalue equation:

$$\sum_{j\beta} \epsilon_{i\alpha,j\beta}(q_\|, \omega) f_{n,j\beta}(q_\|, \omega) = \epsilon_n(q_\|, \omega) f_{n,i\alpha}(q_\|, \omega), \quad (26)$$

which returns the eigenvalues, $\epsilon_n(q_\|, \omega)$, and eigenvectors, $f_{n,i\alpha}(q_\|, \omega)$ of the dielectric matrix in the monopole/dipole basis. A plasmon eigenmode fullfills that:

$$\mathrm{Re} \sum_{j\beta} \epsilon_{i\alpha,j\beta}(q_\|, \omega) f_{n,j\beta}(q_\|, \omega) = 0, \quad (27)$$

corresponding to $\mathrm{Re} \epsilon_n(q_\|, \omega) = 0$. In practice, the plasmon energies are identified from the peaks in the eigenvalue loss-spectrum $-\mathrm{Im} \epsilon_n(q_\|, \omega)$ since this includes the finite imaginary part which can shift the plasmon energy. The right eigenfunctions $f_{n,i\alpha}$ give the induced potential of the plasmon in the basis of $\phi_{i,M/D}$. The left eigenfunctions, $f^n_{i\alpha}$, correspond to the induced density of the plasmon in the basis of $\rho_{i,M/D}$[3]. The induced density is thus given by

$$\rho_n(z, q_\|) = \sum_{i\alpha} f^n_{i\alpha} \rho_{i\alpha}(z, q_\|) \quad (28)$$

with the corresponding induced potential

$$\phi_n(z, q_\|) = \sum_{i\alpha} f^n_{i\alpha} \Phi_{i\alpha}(z, q_\|) \quad (29)$$

## IV. EXCITONS

The Mott-Wannier model, widely used to model excitons in bulk semiconductors, can be straightforwardly generalised to 2D semiconductors. This leads to a 2D hydrogenic Hamiltonian of the form

$$\left[ -\frac{\nabla^2_{2D}}{2\mu_{ex}} + W(\mathbf{r}) \right] F(\mathbf{r}) = E_b F(\mathbf{r}), \quad (30)$$

where $F(\mathbf{r})$ is the exciton wave-function, $\mu_{ex}$ the exciton effective mass and $W(\mathbf{r}_\|)$ is the screened Coulomb potential which includes the screening coming from the 2D material itself and the environment, e.g. a substrate.

Now, consider electron and hole charge distributions given by (the in-plane variation is a plane wave of wave vector $q_\|$)

$$\rho^{e/h}(z, q_\|) = \sum_{i\alpha} \rho^{e/h}_{i\alpha}(q_\|) \rho_{i\alpha}(z, q_\|) \quad (31)$$

We can then calculate the screened interaction between the electron and hole charge distributions according to

$$W(q_\|) = \sum_{k\alpha,i\beta,j\gamma} \rho^e_{k\alpha}(q_\|) \epsilon^{-1}_{k\alpha,i\beta}(q_\|) V_{i\beta,j\gamma}(q_\|) \rho^h_{j\gamma}(q_\|). \quad (32)$$

In the case of excitons located in the layer 1 we can approximate $\rho^{e/h}(z, q_\parallel) = \rho_{1M}(z, q_\parallel)(z)$ and we recover the expression in the Methods section. We can describe a general charge distribution, e.g. using conduction/valence band charge distributions $\rho^{e/h}(z, q_\parallel) = |\psi_{c/v}(z, q_\parallel)|^2$, by a simple redefinition of the Coulomb matrix elements in Eq. 32.

Performing a 2D Fourier transform of $W(q_\parallel)$ yields the screened potential in real space:

$$W(\mathbf{r}_\parallel) = -\frac{1}{2\pi} \int_0^\infty dq_\parallel q_\parallel J_0(q_\parallel \mathbf{r}_\parallel) W(q_\parallel), \quad (33)$$

where $J_n(x)$ is the Bessel function of the first kind. The exciton mass can be obtained e.g. from an ab-initio band structure calculation. We solve Eq. 30 using polar coordinates and a logarithmic radial grid.

## V. SCREENING OF PERPENDICULAR FIELDS

In Fig. 3c in the manuscript, the spatial form of the response due to a constant perturbation across a N=50 layer slab of $MoS_2$ is shown. This gives the dielectric function due to a finite wavevector in the plane of the material. However, the model can also be used to calculate the response to fields with a variation in the $z$−direction, perpendicular to the layers, and can thus be used to calculate the $z$−component of the dielectric function, $\epsilon_{zz}$. This can be calculated in the optical limit, $q_z \to 0$ with the expression:

$$\epsilon_{zz}^{-1} = \frac{12}{L^3} \int\limits_{-L/2}^{L/2} \int\limits_{-\infty}^{\infty} z\,\epsilon^{-1}(z, z'\omega = 0)\, z'\, dz'dz, \quad (34)$$

where $L$ is the width of the structure. In the heterostructure model, this corresponds to taking the matrix product of $\epsilon_{i\alpha,j\beta}^{-1}(q_\parallel, \omega)$ with a vector, $v$, with the elements: $v_{j\beta} = \delta_{\beta, D}$, where only the dipole elements are non-zero: $v = \{0, 1, 0, 1, ...\}$. The expression becomes:

$$\epsilon_{zz}^{-1} = \frac{1}{N} \sum_{i,j} v_{i,D} \, \epsilon_{iD,jD}^{-1}(q_\parallel = 0, \omega = 0)\, v_{j,D}, \quad (35)$$

where $N$ is the number of layers.

In Fig. 2 the induced potential of a N=4 layer $MoS_2$ slab due to an external potential with a linear form along $z$, $V_{ext}(z) \propto z$ is shown together with the ab initio result. The potential is clearly screened by the material, where the induced potential has opposite sign that the external potential. The ab initio result is in this case obtained by applying a weak electric field (within the linear response regime) in the $z$-direction on the ground-state DFT level. This calculation was performed on a real-space grid representation of the electronic wavefunctions, with a grid-spacing of $h = 0.18$ Å, and $(12, 12)$ $k$-points, which were sufficient to converge the ground-state electronic density

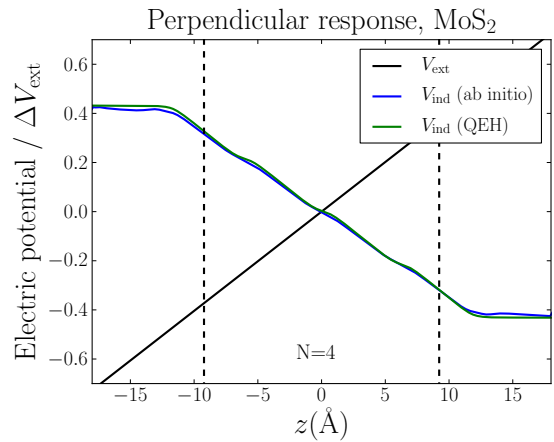

Perpendicular response, $MoS_2$

FIG. 2: Induced potential of a N=4 layer $MoS_2$ slab, due to an external perturbation with a constant slope across the structure. The potentials are normalized with respect to the potential drop across the structure, $\Delta V_{ext}$, the width of the structure here defined as $L = 4d_{MoS_2} = 24.6$Å. The dashed lines indicate the center of the outermost layers.

that determines the total potential. The induced potential is then obtained as $V_{ind} = V_{tot} - V_{ext}$.

As seen in Fig. 2, the QEH model captures the response to perpendicular fields quite well, with a tendency to overestimate the drop in induced potential across the structure and therefore overestimate the dielectric function. This leads to a value of $\epsilon_{zz}(QEH) = 7.71$ compared to an ab initio value of $\epsilon_{zz}(ab\ initio) = 6.81$ for the $N = 4$ $MoS_2$ slab. In case of bulk $MoS_2$ we obtain $\epsilon_{zz}(bulk, ab\ initio) = 6.03$ compared to $\epsilon_{zz}(N = 100, QEH) = 7.83$, which means that the bulk limit is less well-described. However, this is to be expected since the model cannot account for the bulk limit as $q_\parallel \to 0$, since the dielectric function $\epsilon(q_\parallel \to 0) = 1$ for finite slab widths in the model, while for a 3D system the dielectric function tends to a finite value.

## VI. COMPUTATIONAL DETAILS

### A. Multilayer $MoS_2$

Ab initio calculations were performed for monolayer $MoS_2$ to obtain the monolayer density response functions and induced densities used as input for the heterostructure model. The single-particle energies and wave functions were calculated with the PBE exchange correlation functional, with a plane-wave basis set with an energy cutoff of 400 eV. A dense $k$-point sampling of $(128, 128)$ in the 2D Brillouin zone was used in order to calculate the response at low momentum transfers. In the linear response RPA calculation we used an energy cutoff of 50

eV for the reciprocal lattice vectors. We used a nonlinear frequency grid from 0 to 35 eV, with an initial grid spacing of 0.02 eV and a broadening of 0.04 eV. Corresponding ab initio calculations were performed for bulk and bilayer $MoS_2$, but with a $k$-point sampling of $(64, 64, 1)$ for the bilayer and $(64, 64, 8)$ for bulk. For the monolayer and bilayer calculations the truncated Coulomb kernel, see Eq. 10, was used while the full, i.e. non-truncated kernel, was used for the bulk calculation. We used an in-plane lattice constants of 3.18 Å, and A-B stacking with 6.15 Å separation between layers. For the monolayer and bilayer calculation the unit cells contained 20 Å of vacuum to separate the periodic images in the $z$-direction. For the heterostructure calculation we used the same separation between the layers as for the ab initio calculations ($d = 6.15$ Å). We note that the effect of stacking arrangement (A-A or A-B) cannot be accounted for within the model.

### B. Graphene/hBN heterostructures

Ab initio calculations were performed to obtain the dielectric building blocks of monolayer doped graphene and hBN. Also, full ab initio calculations were done for entire heterostructures, including up to three layers of hBN, or the equivalent amount of vacuum, separating the doped graphene layers. An in plane lattice-constant of 2.5 Å was used for both graphene and hBN, so that the heterostructure could be represented a $1 \times 1$ unit cell. The layers were stacked in A-B configuration, with 3.326 Å separation ($c$-lattice constant of 6.653). We used PBE exchange-correlation, a 340 eV energy cutoff for the plane waves in the ground state calculations, and $(100, 100)$ $k$-point sampling in the 2D Brillouin zone. In the response calculation doped structures were obtained by shifting the Fermi-level 1 eV upwards. An energy cutoff of 70 eV was used for the reciprocal lattice vectors, and unoccupied bands were included up to 35 eV above the Fermi level. All the calculations employed the truncated Coulomb interaction and 20 Å vacuum to separate the repeated structures. A non-linear frequency-grid with an initial grid spacing of 0.02 eV and a broadening of 0.05 eV was used to represent the dynamic response function. Plasmon eigenmodes were obtained by diagonalizing the dielectric matrix in Bloch representation as described in ref.[3].

### C. Excitons in supported $WS_2$

The dielectric building blocks of the $WS_2$, hBN, and $MoS_2$ monolayers were calculated as follows. Single-particle energies and wave functions were calculated using LDA, a plane wave cut-off of 500 eV, and $(45, 45)$ $k$-points. The density response function was calculated within RPA using an energy cut-off of 300 eV and including empty states up to 50 eV above the Fermi level. The truncated Coulomb kernel was employed and 20 Å vacuum was included in the supercell to separate repeated layers. In setting up the heterostructure we used a separation of 3.22 Å between the 100 layers of h-BN and 5.08 Å between $WS_2$ and h-BN. For $WS_2$ on 50 layers of $MoS_2$ we used a uniform separation of 6.3 Å between all layers. We then calculated the screened interaction from Eq. 32 for $q_\parallel$ up to (and including) the second Brillouin zone. For calculating the exciton Rydberg series we solved Eq. 30 for spherical states on a radial logarithmic grid and verified that the exciton energies were converged to within 0.01 eV.

### D. 2D Database

The dielectric building blocks were calculated for 51 transition metal dichalcogenides and oxides, hBN, and graphene at 10 different doping levels from 0.1 to 1 eV. For the single particle wave functions and energies obtained from DFT, we used PBE exchange-correlation and a plane-wave basis with a energy cutoff equal to 500 eV. The 2D Brillouin zone was sampled by $(200,200)$ $k$-points for graphene, and for the remaining materials we used a $k$-point density corresponding to $(100,100)$ $k$-points.

For the density response functions we used a cutoff of 100 eV for the transition metal dichalcogenides and oxides and 150 eV for graphene and hBN. The truncated Coulomb kernel was employed and 20 Å vacuum was included in the supercell to separate the repeated layers. All materials were represented on the same frequency grid from 0 to 35 eV, with an initial spacing of 0.01 eV and a broadening of 0.05 eV. The response functions were calculated for a range of in-plane momentum transfers, $q_\parallel$, within the first Brillouin zone of graphene up to a maximum value of $q_\parallel = 2.89$ Å$^{-1}$. At small $q_\parallel$ below 0.3 Å$^{-1}$ we use a denser sampling with a grid spacing of 0.015 Å$^{-1}$ in order to capture the strong $q_\parallel$-dependence of the plasmon energies and the dielectric function in this region. After this limit the grid spacing is increased to 0.029 Å$^{-1}$. In order to obtain all response functions on the same $q_\parallel$-grid, the data for the remaining materials was interpolated to the grid for graphene using conventional 2D spline interpolation.

---

* Electronic address: thygesen@fysik.dtu.dk


\end{document}